\documentclass[aps,prl,twocolumn,groupedaddress,floatfix,longbibliography]{revtex4-1}
\usepackage{graphicx,color}  % needed for figures
\usepackage{dcolumn}   % needed for some tables
\usepackage{bm}        % for math
\usepackage{amssymb}   % for math

\usepackage{ulem}
\usepackage[english]{babel} 
\usepackage{amssymb,amsfonts,amsmath}
\usepackage{graphicx,bm}
\usepackage{color}
\usepackage{epsfig,dsfont,amssymb,amsmath,amsthm,amsfonts,amsbsy,mathrsfs}
\usepackage{color}
\usepackage{xcolor}

\usepackage{tikz}

\definecolor{red}{rgb}{1,0,0}
\definecolor{blue}{rgb}{0,0,1}
\definecolor{black}{rgb}{0,0,0}

\usepackage{amsmath}
\begin{document}

\title{Mean-field description of plastic flow in amorphous solids}
\author{Jie Lin}
\affiliation{Department of Physics, New York University, New York, New York 10003, USA}
\author{Matthieu Wyart}
\affiliation{Institute of Theoretical Physics, Ecole Polytechnique Federale de Lausanne (EPFL), CH-1015 Lausanne, Switzerland}

\date{\today}
\begin{abstract}
Failure and flow of amorphous materials are central to various phenomena including earthquakes and landslides. 
There is accumulating evidence that the yielding transition between a flowing and an arrested phase is a critical phenomenon, 
but the associated exponents are not understood, even at a mean-field level where the validity of popular models is debated. 
Here we solve a mean-field model that captures the broad distribution of the mechanical noise generated by plasticity,  
whose behavior is related to biased L\'evy flights near an absorbing boundary. We compute the exponent $\theta$ characterising the density of  shear transformation $P(x)\sim x^{\theta}$, where $x$ is the stress increment beyond which they yield. We find that after an isotropic thermal quench, $\theta=1/2$. However, $\theta$ depends continuously on the applied shear stress, this dependence is not monotonic, and its value at the yield stress is not universal. The model rationalizes previously unexplained observations, and captures reasonably well the value of exponents in three dimensions. Values of exponents in four dimensions are accurately predicted. These results support that it is the true mean-field model that applies in large dimension, and raise fundamental questions on the nature of the yielding transition.
\end{abstract}
\pacs{}
\maketitle
\section{Introduction}
Amorphous solids such as emulsions, glasses or sands are yield stress materials, which fail and flow if a sufficient shear stress is applied.
In the solid phase, plasticity can be conceived as consisting of  elementary rearrangements, the so-called shear transformations \cite{Argon79,Falk98,Schall2007,Amon2012,Tanguy06}. Shear transformations are localized but display long-range elastic interactions \cite{Picard04}, and organize dynamically into elongated  highly plastic regions\cite{Amon2012,Lebouil2014,Gimbert2013,Lemaitre09,Maloney09}. Above some threshold stress, failure  occurs and one enters a fluid phase where a stationary flow can be maintained. In various materials, rheological properties  appear to be controlled by a critical point at the yield stress $\Sigma_c$ where the flow stops: at that point, flow curves relating shear stress and strain rate are singular  \cite{Bonn15}, and the dynamics displays long-range spatial correlations \cite{Martens11,Salerno12,Lemaitre09}. Despite the importance of these properties in a variety of phenomena including earthquakes and landslides, a quantitative  microscopic description is lacking. As is generally the case in condensed matter systems, one expects that the density of elementary excitations strongly affects such properties. For amorphous solids this corresponds to the density $P(x)$ of shear transformations, characterized by the additional shear stress $x$ required to trigger them \cite{Lemaitre07,Karmakar10a}. One finds empirically a {\it pseudo-gap}, i.e. $P(x)\sim x^\theta$, for small $x$ with $\theta>0$\cite{Lin14a,Salerno13,Karmakar10a}. The value of $\theta$ was argued to control the singular rheological properties and diverging length scale of flow just above the yield stress $\Sigma_c$ \cite{Lin14}. The fact that $\theta>0$ was also shown to imply crackling (system spanning avalanche-type response) in the entire solid phase $\Sigma<\Sigma_c$\cite{Lin15},  where $\Sigma$ is the applied shear stress, and the value of $\theta$ affects avalanche statistics in that regime. 

Pseudo-gaps are commonly found in glassy systems with sufficiently long-range interactions \cite{Muller14}. The associated exponent is constrained by stability, as occurs in  electron glasses \cite{Efros75}, fully-connected spin glasses\cite{Thouless77,Pazmandi99,Doussal10,Eastham06,Yan15} and hard sphere packings \cite{Wyart12,Lerner13a,Kallus13,Charbonneau14,Charbonneau15}. In the last two cases, the associated stability bound can be proven to be saturated (a scenario referred to as marginal stability) \cite{Muller14}. For amorphous solids, given the elastic coupling between shear transformations, stability toward extensive avalanches can be shown to imply $\theta>0$ \cite{Lin14a}. However, most recent data indicate that (i) for quasi-static flows (at the yield stress) $\theta\in [0.5, 0.57]$ and $\theta\in [0.35,0.43]$ in two and three dimensions respectively \cite{Lin14,Salerno13}, (ii) right after a fast quench at zero stress, $\theta\in [0.4, 0.6]$ \cite{Lin14a,Karmakar10a} and (iii) as the stress increases within the solid phase, $\theta$ rapidly drops initially, and then slowly rises again  as $\Sigma_c$ is approached from below \cite{Lin15}. Thus the marginal stability scenario does not yield the pseudo-gap exponent value for amorphous solids, and moreover, cannot explain its non-monotonic dependence on the applied stress.

An alternative route seeks progress by considering mean-field models, that would allow one to compute $\theta$ in large spatial dimension $d$, where spatial correlations between local plastic events are presumably weak. Hebraud and Lequeux (HL)\cite{Hebraud98} introduced a popular model where all shear transformations interact with each other, with a similar magnitude. A pseudo-gap is predicted, but one finds $\theta=1$ which is far from the values observed in two and three dimensions, and does not depend on the applied stress. Lemaitre and Caroli (LC) pointed out that since elastic interactions decays as an inverse power of distance, the magnitude between two randomly chosen shear transformations is broadly distributed \cite{Lemaitre07}. Including this effect led to a mean-field model which was numerically shown to display a pseudo-gap, but not solved analytically. 

We introduce a class of mean-field models that interpolate continuously between these two cases, and solve them using a combination of probabilistic arguments and analysis. In our models, spatial correlations are neglected.  However, the distribution of stress fluctuations generated by a local event is kept the same as finite dimensional systems. Because of the broad distribution of mechanical noises, the variables $x$ describing the stability of shear transformations undergo biased one-dimensional L\'evy-flights of index $\mu$ with absorbing conditions outside a compact interval, and re-insertion within this interval. HL model corresponds to $\mu\geq 2$ (Brownian motion), whereas the more physical LC model corresponds to $\mu=1$. Our findings are that (a) for $\mu>1$, $\theta$ is independent of system preparation and follows $\theta=\mu/2$ for $\mu\in (1,2]$. (b) For $\mu<1$, $\theta=\mu/2$ after an isotropic ($\Sigma=0$) quench but $\theta=0$ if $\Sigma>0$, in particular at the yield stress $\Sigma_c$. (c) For the physical case $\mu=1$, $\theta=1/2$ after a quench, but $\theta$ is {\it not universal} for $\Sigma>0$, and is shown to drop immediately as $\Sigma$ increases from zero, and then increases continuously with the applied stress. These predictions are confirmed numerically. They are remarkably consistent with finite-dimensional observations that were unexplained even at a qualitative level, in particular regarding the non-monotonicity of the pseudo-gap exponent with the applied shear stress. Quantitatively, the values we predict for exponents are already reasonably accurate in three dimensions, and become very precise in four dimensions. These fact support that our mean-field model is the true mean-field model that applies in high spatial dimensions. A surprising consequence of our approach is that the non-universal value $\theta(\Sigma_c)$ will never reach a well-defined value above some critical dimension. Instead, it simply tends to decrease with $d$, in agreement with observations. 

Beyond plasticity, to our knowledge our model provides the first non-trivial glassy system where the violation of marginal stability can be proven ({\it i.e.}, the fact that the pseudo-gap exponent is strictly larger than what required by stability). Another significant byproduct of our work is the classification of the asymptotic scaling behavior of a biased L\'evy flight close to an absorbing boundary.

Sec.\ref{MFD} introduces mean field models. Their thermodynamics limits are worked out  in Sec.\ref{CD}, together with a derivation of the pseudo gap exponent. A physical interpretation of these results based on survival probability of biased Levy Flights is presented in Sec.\ref{IN}. Predictions are tested numerically  in Sec.\ref{NT}. History-dependence of the pseudo-gap exponents  is studied in Sec.\ref{TB}, while Sec.\ref{RSD} investigates the role of spatial dimension. We conclude by summarizing the consequences of our work for real materials and by raising open questions.

\begin{figure}[hbt!]
\centering \includegraphics[width=.48\textwidth]{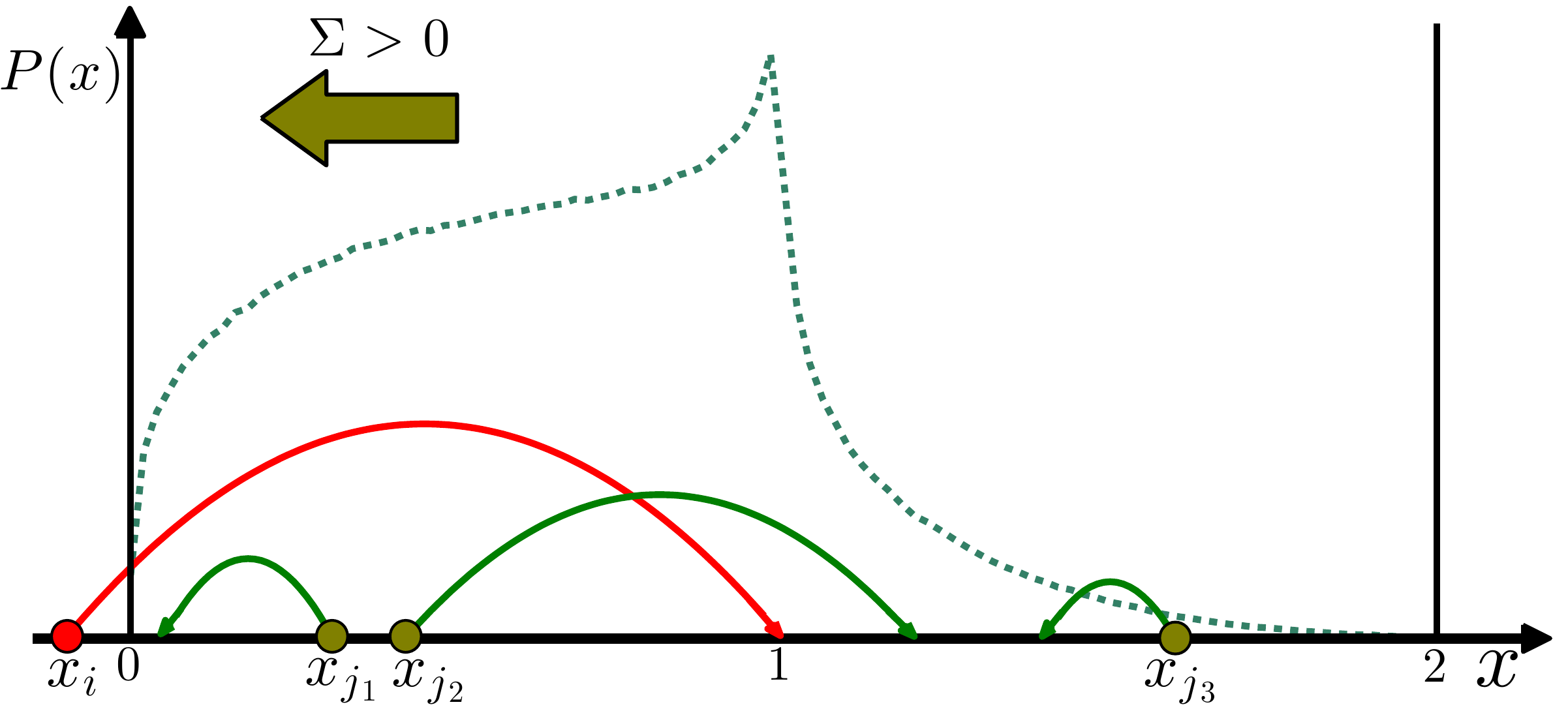} 
\caption{Mean field model: an unstable site (red line) returns to $x_i=1$ from $x_i<0$ after a time $\tau_r$. Concomitantly, other sites implement a drift toward negative $x$ to conserve the mean stress, and a random jump of symmetric distribution $w(\xi)$. The dashed line represents $P(x)$, assuming $\Sigma>0$. A  pseudo-gap is represented at $x=0$. }\label{cartoon}
\end{figure}

\section{\label{MFD}Mean Field Models}
Following \cite{Baret02,Picard2005,Martens11,Talamali11,Lin14a}, we describe amorphous materials as consisting of $N$ blocks, each characterized by a local shear stress $\sigma_i$ and a local yield stress $\sigma_i^{th}$, which we choose to be unity. The shear stress applied on the material is $\Sigma=\sum_i \sigma_i/N$, and we assume that $\Sigma$ is constant in time (we relax this hypothesis below).  A block becomes unstable if $|\sigma_i|>\sigma^{th}\equiv 1$. This condition is easily expressed by defining  $x_i\equiv1-\sigma_i$ and stability corresponds to $x_i\in [0,2]$. If the variable $x_i$ exits this interval at some time $t$, the block $i$ is declared ``unstable".  After some constant time interval $\tau_r$ (describing the time for a local plastic event to occur) $\sigma_i$ relaxes to zero, {\it i.e.,} $x_i\rightarrow 1$ at time $t+\tau_r$ (allowing for a distribution of stress drop amplitude does not affects our results below). For this choice of dynamics, $x_i\notin [0,2]$ are always absorbing conditions for the variable $x_i$. (Another popular choice of dynamics assumes that if $x_i\notin [0,2]$, there is a finite evaporation rate at which sites become unstable. Such a dynamic  leads to an absorbing condition outside $[0,2]$ only in the quasi-static limit. For such models our results below must apply in that limit, as supported by the fact that the pseudo-gap exponent does not depend on the choice of dynamics then \cite{Lin14a}). 

Such a relaxation event corresponds to a plastic strain in site $i$ of magnitude $\Delta\epsilon_i= \sigma_i /E$, where $E$ is the shear modulus. This relaxation causes a global plastic strain $\Delta\epsilon=\Delta\epsilon_i/N$. Moreover, the stress in $i$ is redistributed on other sites, leading to:
\begin{equation}
\label{00}
x_j(l+1)=x_j(l)+\mathcal{G}_{ji}(\vec{r}_i-\vec{r}_j) \sigma_i (l)
\end{equation}
where $\mathcal{G}_{ji}(\vec{r}_i-\vec{r}_j)$ is the interaction kernel which a priori depends on the position $\vec{r}_j$ of the sites, and the integer $l$  numbers  plastic events in chronological order.  If the stress is fixed one must have $\sum_{j\neq i} {\cal G}_{ij}=-{\cal G}_{ii}=-1$.  In the following, we set $E=\tau_r=1$.

For amorphous solids, ${\cal G}(r)$ is well-approximated by the Eshelby kernel, of magnitude ${\cal G}(r)\sim 1/r^d$ and whose sign depends on the relative directions between $(\vec{r}_i-\vec{r}_j)$ and the imposed shear  \cite{Picard04,Kabla03, Desmond13}. This property implies that the distribution $\rho(\Delta x)$ of kicks $\Delta x_j=x_j(l+1)-x_j(l)$ at each plastic event is broadly distributed, since site $i$ can be close or far from site $j$. Using ${\cal G}(r)\sim 1/r^d$, one readily finds \cite{Lemaitre07} that $\rho(\Delta x)\sim |\Delta x|^{-1-\mu}$ with $\mu=1$. It is straightforward to extend this result to the general case ${\cal G}(r)\sim 1/r^\alpha$, where we find $\mu=d/\alpha$. Extending \cite{Lemaitre07}, mean field models can now be constructed where the distribution of  kicks amplitudes preserves that of the original problem, but where all sites interact statistically in an identical way. This is achieved by replacing the prescription Eq.(\ref{00}) for the relaxation of site $i$ by the new rule:
\begin{align}\label{dynamics}
x_i(l+1)&= 1 \nonumber \\ 
x_j(l+1)-x_j(l)&= -\frac{1-x_i(l)}{N} +\xi_j 
\end{align}
In the last equation, the first term on the right side (referred to as drift term below) ensures conservation of stress, {\it i.e.}, $\sum_i x_i(l)$ does not depend on $l$. The random variable $\xi$  has zero mean $\langle \xi_i \rangle=0$, and its distribution $w(\xi)$ mimics  that of the finite dimensional model:
\begin{equation}
w(\xi)=\frac{A}{N} |\xi|^{-\mu-1} \label{wx},
\end{equation}
with a lower cut-off  $\xi_{c}=(\frac{2A}{\mu})^{1/\mu}N^{-1/\mu}$ fixed by normalization  and an upper cut-off $\xi_{m}=(\frac{2A}{\mu})^{1/\mu}$ (such a cut-off is present in the finite-dimensional model, and corresponds to the amplitude of the kick given by an adjacent site). The dynamical rule of the mean field models are illustrated in Fig.\ref{cartoon}.%} {\blue Without the random kicks, the present model is equivalent to the mean field depinning model, where $\theta$ is trivially $0$\cite{Fisher98}.}

Such mean field models behave qualitatively like standard elastoplastic models: there exists a yield stress $\Sigma_c>0$ such that for $\Sigma<\Sigma_c$, the dynamics eventually stops, corresponding to the solid phase. For $\Sigma>\Sigma_c$, the dynamics never stops in the thermodynamic limit, and is characterized by a  rate of plastic strain $\dot{\epsilon}=N_u\langle \sigma_u\rangle/N$, where $N_{u}$ is the instantaneous number of unstable blocks, and $\langle \sigma_u\rangle=1-\langle x_u\rangle$ their mean stress value when they relax. $\langle \sigma_u\rangle$ has both a positive contribution from $\sigma_u>1$ ({\it i.e.} $x_u<0$) and a negative one from $\sigma_u<-1$ ({\it i.e.} $x_u>2$), and the symmetry between these is broken as soon as $\Sigma>0$. In our convention, for $\Sigma>0$, $\langle x\rangle<1$ and most sites become unstable at the boundary  $x<0$, leading to $\dot\epsilon>0$. 
 
Ultimately, our model  describes L\'evy flights with absorbing conditions for $x\notin [0,2]$. Due to the drift term in Eq.(\ref{dynamics}) these flights are biased, which tends to bring them toward the unstable region $x<0$ if $\Sigma>0$. For $\Sigma>\Sigma_c$ where a stationary state is reached, computing the pseudo-gap in this mean-field approximation requires to obtain  the stationary distribution of the stable sites $P(x)\equiv \sum_{stable} \delta(x-x_i)/N$ of biased L\'evy flights near an absorbing boundary. %In our notation, only the stable sites enters in $P$, their fraction $1-N_u/N$ goes to one however at $\Sigma_c$ where $\dot{\epsilon}\rightarrow 0$.

\begin{table}[!htb]
\caption{\small{Summary of results: mean field values (MF) of $\theta$ at the yield stress $\Sigma_c$ (implying $v>0$) and after a quench at $\Sigma=0$ (for which $v=0$) as a function of the L\'evy index $\mu$, the random kick amplitude $A$, and the bias $v$. For comparison we also report $\theta$ (marginality) corresponding to the saturation of the stability bound derived in \cite{Lin14a}.}}\label{table1}
\begin{tabular}{ c | c | c | c }
\hline
  $\mu$             &  $\theta$($\Sigma_c$)   & $\theta(\Sigma=0)$&  $\theta$ (Marginality) \\ [3pt] \hline
  $ \mu\geq 2$                                  &  1                                      &     1      &1                 \\[5pt] \hline
  $ 1<\mu<2$                        &   $\mu/2$                         & $\mu/2$ & $\mu-1$     \\[5pt] \hline
  $\mu=1$                                        &\normalsize $\arctan(\pi A/v)/\pi$  & $1/2$& 0   \\[5pt] \hline
  $\mu<1$                                      &    0                                     &    $\mu/2 $ &  0    \\[3pt] \hline  
\end{tabular}
\end{table}
\section{\label{CD}Continuous description}
We consider the  limit $N\rightarrow \infty$ while keeping the variable $\gamma\equiv l/N$ fixed, with $\gamma\ll 1$ ($\gamma$ is essentially a measure of the accumulated plastic strain $\epsilon$, as $\epsilon=\gamma \langle \sigma_u\rangle$ where in practice $\langle \sigma_u\rangle\simeq 1$  for $\Sigma\simeq \Sigma_c$). In this limit, the dynamics of stable sites in Eq.(\ref{dynamics}) becomes:
\begin{equation}
\label{02}
x_j(\gamma)=x_j(0) - v\gamma +\xi_j(\gamma)
\end{equation} 
Here the drift follows:
\begin{equation}
v= \langle \sigma_u \rangle =1-\langle x_u\rangle 
\end{equation} 
For this convention  $v>0$ if $\Sigma>0$. We assume that $v>0$ and will relax this hypothesis when discussing thermal quenches at $\Sigma=0$. The random kick $\xi_j(\gamma)$ is an accumulation of $\gamma N$ discrete random kicks, $\xi_j(\gamma)=\sum_{k=1}^{\gamma N}\xi_k$, and satisfies the probability distribution 
\begin{equation}
w_{\gamma}(\xi)=\int    \delta(\sum_{k=1}^{\gamma N}\xi_k-\xi)  \prod_{k=1}^{\gamma N}w(\xi_k)d\xi_k
\end{equation}
In Fourier space, $\tilde{w}_{\gamma}(k)=\tilde{w}(k)^{\gamma N}$, where $\tilde{w}(k)$ is the Fourier transform of $w(\xi)$ defined in Eq.(\ref{wx}). 

According to Eq.(\ref{02}), together with the rule that unstable sites are reinserted in $x=1$, one obtains the time evolution of the density distribution of $x$, $P(x)$, from $\gamma$ to $\gamma+\delta\gamma$:
 \begin{align}
&P(x,\gamma	+\delta\gamma )=P(x+v\delta\gamma,\gamma)\\
&+\int_{-\infty}^{\infty} P(y,\gamma) w_{\delta\gamma}(y-x)dy-P(x,\gamma)\int_{-\infty}^{\infty} w_{\delta\gamma}(y)dy\nonumber\\
&+\delta\gamma\delta(x-1)\nonumber+O(\delta \gamma^2)\nonumber
\end{align}
where the first term on the right side represents the drift, the second term characterizes the flux of particle arriving in $x$ and the third term  the flux of particles departing from $x$. Eventually, we obtain the  master equation:
\begin{align}
\label{master}
\frac{\partial P}{\partial \gamma}&=v\frac{\partial P(x)}{\partial x } + \int_{-\infty}^{\infty} [P(y)-P(x)]w^{\prime}(y-x) dy + \delta(x-1)
\end{align}
with the condition that $P(x)=0$ if $x\notin [0,2]$, and where  $w^{\prime}(\xi)=\lim_{\gamma\rightarrow 0 }w_{\gamma}(\xi)/\gamma$. 

\begin{figure}[hbt!]
\centering\includegraphics[width=.36\textwidth]{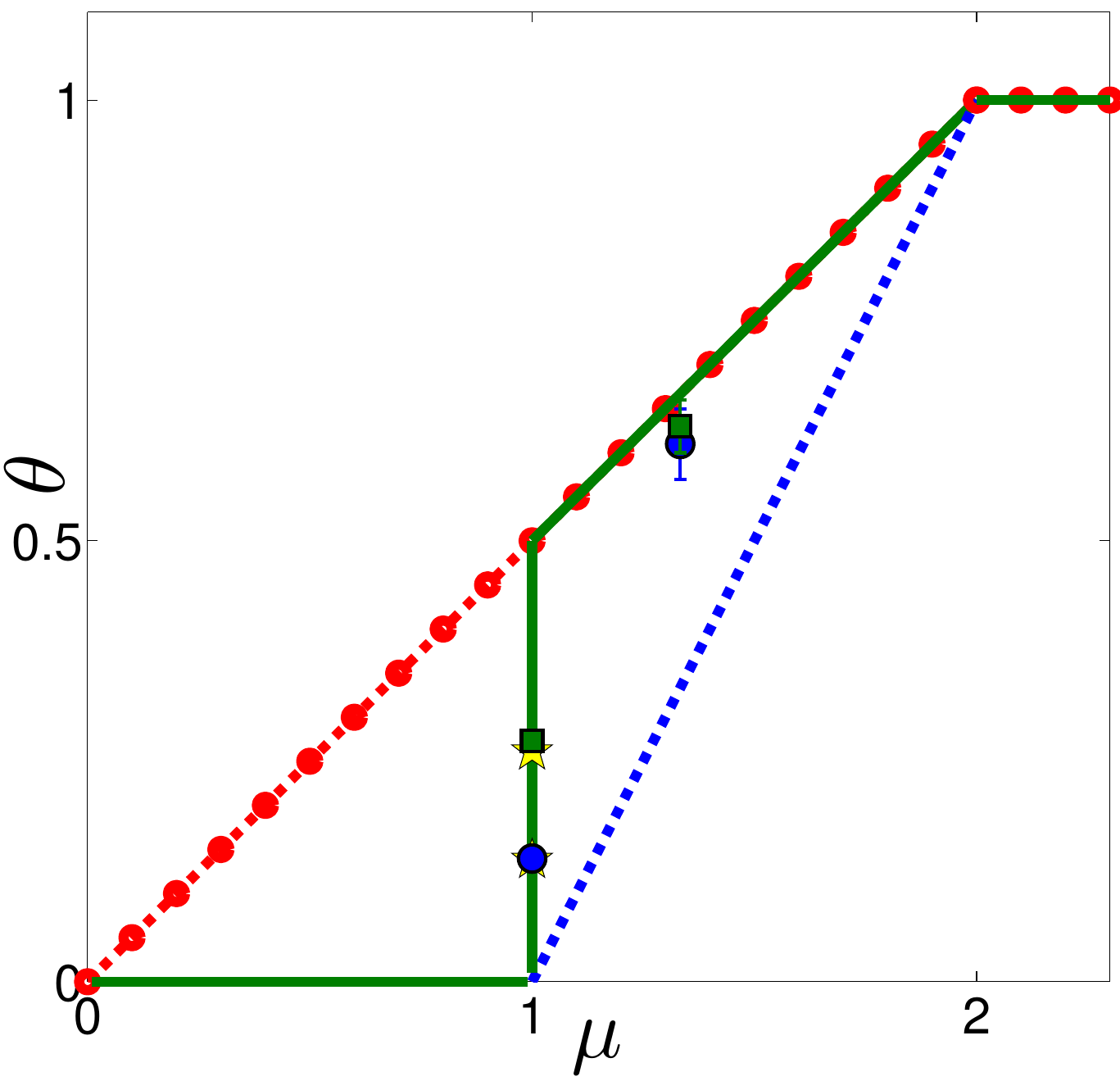} 
   \caption{Theoretical prediction of  the pseudo gap exponent $\theta$ {\it v.s} the L\'evy index of the random kicks $\mu$. Green line: $\theta$ at  the yield stress $\Sigma_c$; Red-dotted line: $\theta$ after a quench, $\Sigma=0$, overlapping with the green line for $\mu>1$; Blue dashed line: marginal values of $\theta$, overlapping with the green line for $\mu>2$, and $0<\mu<1$. Below the blue line, the system is unstable and forbidden dynamically \cite{Lin14a}. The data points are the measured value of $\theta$ at $\Sigma_c$ in the simulated mean-field-model: for $\mu=4/3$, $A=0.15$ (circle), $A=0.35$ (square); $\mu=1$, $A=0.15$ (circle), $A=0.35$ (square), and the yellow stars are the corresponding theoretical values indicated when $\mu=1$.}\label{figure1}
\end{figure}

{\bf Case $\mu\geq 2$:}  $w(\xi)$ has then a finite variance, and $\tilde{w}(k)=1-\langle \xi^2\rangle k^2/2+O(k^4)$.  For $\tilde{w}_{\gamma}(k)$ to converge in the large $N$ limit, one must choose $\langle \xi^2\rangle= 2D/N$ where $D$ is a constant (implying $A\sim N^{1-\frac{\mu}{2}}$), so $\tilde{w}_{\gamma}(k)\rightarrow \exp(-\gamma Dk^2)$. $w^{\prime}(\xi)$ then converges to a Gaussian, and Eq.(\ref{master}) leads to the standard Fokker-Planck equation:
\begin{equation}
\frac{\partial P}{\partial \gamma}=v \frac{\partial P}{\partial x} + D\frac{\partial^2 P}{\partial x^2} +\delta (x-1)
\end{equation}
Solutions of such a diffusion equation vanish linearly near the absorbing condition, {\it i.e.}, $P(x)\sim x$ at small $x$, as found in the HL model \cite{Hebraud98}. This corresponds to $\theta=1$

{\bf Case $\mu<2$:} In that case, one recovers the well-known results for L\'evy distribution \cite{Bouchaud90}:
\begin{equation}
w^{\prime}(\xi)\rightarrow \frac{ A}{|\xi|^{\mu+1}}
\end{equation}
with an upper cut-off at $\xi_{m}$, see Appendix.B for details, Eq.(\ref{master}) then leads to:
\begin{align}
\frac{\partial P}{\partial \gamma}=&v\frac{\partial P(x)}{\partial x}- P(x)\int_{-\infty}^{0}  w^{\prime}(y-x) dy   \nonumber\\
&+\int_0^{2} (P(y) -  P(x)) w^{\prime}(y-x) dy  +\delta(x-1) \label{raweq}
\end{align}
Using the condition that $P(x)=0$ for $x<0$ and $x>2$.
\section{\label{AS}Asymptotic solutions}
We seek stationary solutions of  Eq.(\ref{raweq}) of the  form $P(x)=P_0+C_0 x^{n}$ for $ x\ll 1$, with $n>0$. Defining $s=y/x$, Eq.(\ref{raweq}) reduces to:
\begin{align}
&v n C_0 x^{n-1} - A\frac{P_0+C_0 x^{n}}{\mu} x^{-\mu} \notag \\
&+ x^{n+1} \int_0^{2/x} \{\frac{P(sx)}{x^n}- \frac{P_0}{x^n}-C_0  \}  w^{\prime}(x(s-1))ds =0\label{maineq}
\end{align}
Here we change variable $s=yx$, and denote the last term $T_3$. Using the fact that $w^{\prime}(\xi)=A/|\xi|^{\mu+1}$, in the limit $x\ll 1$, $T_3$ converges to:
\begin{equation}
T_3=AC_0 x^{n-\mu} \int_{0}^{\infty} \frac{s^n-1}{|s-1|^{\mu+1}}ds
\end{equation}
if $n<\mu$, or $T_3=O(1)$ if $n>\mu$.\\

\textbf{Case $\mu<1$}: In this situation we must have $P_0>0$, otherwise the first term in Eq.(\ref{maineq}) cannot be balanced. So $T_3$ is always negligible compared with the first two terms, in both cases $n<\mu$ or $n>\mu$. Keeping only the dominant terms, we obtain:
\begin{equation}
v n C_0 x^{n-1}  -  A \frac{P_0}{\mu} x^{-\mu} =0\label{balance}
\end{equation}
implying $n-1=-\mu$. Thus we find:
\begin{equation}
P(x)=P_0+C_0 x^{1-\mu}\label{case1}
\end{equation}
corresponding to $\theta=0$. 

\textbf{Case  $1<\mu<2$}:
In this case, if $P_0>0$, Eq.(\ref{maineq}) cannot be satisfied because the term proportional to $P_0x^{-\mu}$ cannot be balanced. So $P_0=0$ and $T_3$ is therefore not negligible.
Equating the dominant terms leads to: 
\begin{equation}
  - A\frac{C_0}{\mu} x^{n-\mu} + T_3=0
\end{equation}
If $n>\mu$, the first terms tends to 0, while the second term remains $O(1)$. Thus  $n\leq \mu$, and:
\begin{equation}
\label{003}
\frac{1}{\mu} - \int_{0}^{\infty} \frac{s^{n}-1}{|1-s|^{1+\mu}} ds=0
\end{equation}
We checked that the unique solution of this equation is $n=\mu/2$, a result which has a simple probabilistic interpretation, as discussed in section IV. Thus $P(x)\sim x^{\mu/2}$ and $\theta=\mu/2$.

\textbf{Case  $\mu=1$}:
The most important case physically is also the richest. Solution can exist only if $P_0=0$ and $n<1$, and Eq.(\ref{maineq}) implies asymptotically: 
\begin{equation}
v n x^{n-1}- Ax^{n-1} + Ax^{n-1} \int_{0}^{\infty} \frac{s^{n}-1}{|1-s|^{2}} ds=0
\end{equation}
The last integral yields $I_1=\int_{0}^{\infty}\frac{s^n-1}{|1-s|^2} ds= 1-\pi n\cot(\pi n)$, from which we obtain $v=A\pi\cot(\pi n)$. Thus $P(x)\sim x^{\theta}$, with
\begin{equation}
\theta=\frac{1}{\pi}\arctan(\frac{\pi A}{v})\label{thetaonA}
\end{equation}
implying that $\theta$ continuously depends on the drift $v$ and the magnitude of the noise $A$.\\

The above results are summarized in Table.\ref{table1} and Fig.\ref{figure1}.
\begin{figure}[htb!]
\centering\includegraphics[width=.48\textwidth]{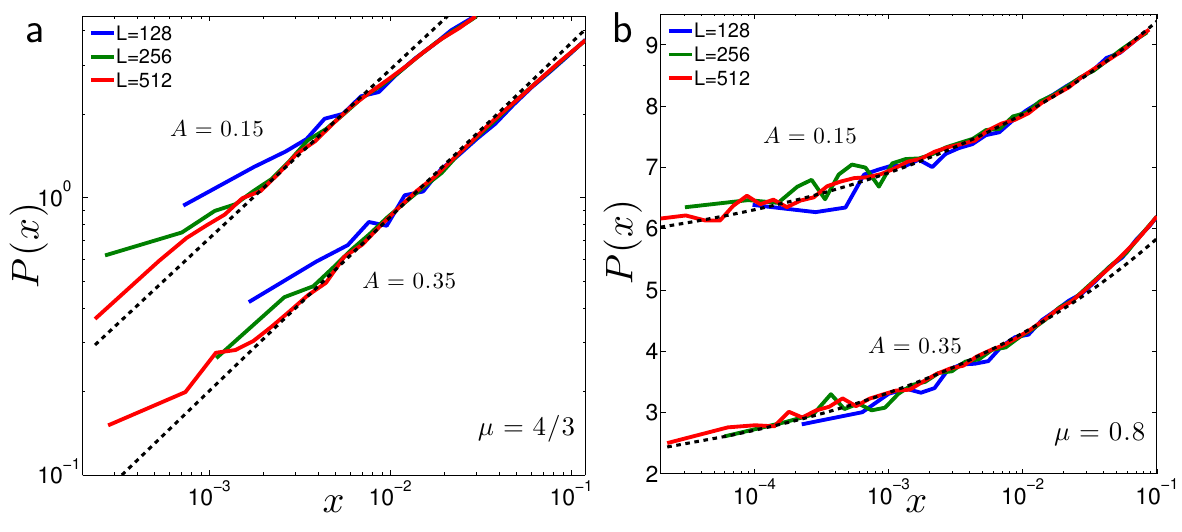} 
    \caption{ $P(x)$ for $\mu=4/3$ (a) and  $\mu=0.8$ (b) for $N=L^2$ as shown in legend. (a) The dashed line is the theoretical value $\theta=0.667$, the measured values are $\theta=0.61\pm 0.04$ ($A=0.15$), $\theta=0.62\pm0.03$ ($A=0.35$). (b) The dased line is the fit of Eq.(\ref{case1}), which are $P(x)=5.273+6.506 x^{0.2}$ ($A=0.15$), $P(x)=1.655+6.618x^{0.2}$ ($A=0.35$).}\label{figure2}
\end{figure}

\section{\label{IN}Interpretation}
A detailed probabilistic derivation of these results is given in Appendix.A, including the case with no bias ($v=0$). In a nutshell, $P(x)$ for $x\ll1$ is proportional to the number of random walkers starting from $x=1$ which ends up in position $x$ after a time $\gamma=O(1)$, without having crossed the absorbing condition $x<0$. For a Brownian motion (corresponding to $\mu\geq2$ in our model) it is well known that this number vanishes linearly in $x$. This result is independent of the bias for a Brownian motion, because on small length scales of order $x$ (or equivalently on small time scale of order $\gamma\sim x^2$), fluctuations always dominate the bias, which is therefore irrelevant. Fluctuations also dominate the bias for Levy flights if $\mu>1$, and thus $P(x)$ at small $x$ must again be independent of the bias, in agreement with our result that $\theta=\mu/2$ if $\mu\in (1,2]$. The case $\mu=1$ however is marginal: bias and fluctuations are always comparable on all time scales, and both affects the value of $\theta$ as shown in Eq.(\ref{thetaonA}). Finally, for Levy flights with $\mu<1$, the bias dominates on small time scales: typically walkers are essentially convected toward the origin, leading to a non-vanishing $P(x)$ at $x=0$.
% {\blue We summarize the mean field predictions of the pseudo-gap exponent and compare them with the marginal bound derived in \cite{Lin14a} in Fig.\ref{figure1}.}

\section{\label{NT}Numerical Tests}
We now test our predictions numerically, considering first the case $\Sigma=\Sigma_c>0$, implying $v>0$. To compute $P(x)$ at $\Sigma_c$ efficiently,  we use the extremal dynamics method, see {\it e.g.} \cite{Talamali11}. Starting from a small value of  stress, the shear stress is increased each time the dynamics stops ({\it i.e.} when there are no more unstable sites).  We choose the stress increment $\delta \Sigma$ to be just sufficiently to trigger a new avalanche of plasticity, {\it i.e.}, $\delta \Sigma=\min\{x_i\}$. During avalanches, the shear stress $\Sigma$ is lowered. This is achieved in practice by removing the  bias in Eq.(\ref{dynamics}), so that the stress drop is of approximatively $1/N$ at each plastic event. Using such dynamics the system spontaneously reaches the stationary state where $\langle \Sigma\rangle =\Sigma_c$. Fluctuations of stress vanish in the thermodynamic limit, and the trajectory of each site is equivalent to that in the fixed stress protocol at the critical stress $\Sigma_c$: both are biased L\'evy flights towards the absorbing boundary with the same drift $v$. We expect our fixed stress predictions to hold, as we confirm numerically. 
Fig.\ref{figure2} shows $P(x)$ for $\mu=0.8$ and $\mu=4/3$ for two different choices of kick amplitude $A$. For $\mu=4/3$,  we measure $\theta$ by fitting the part of the curves that overlap for different system sizes, and find $\theta=0.61\pm 0.04$ ($A=0.15$) and $\theta=0.63\pm 0.03$ ($A=0.35$). These results are slightly smaller but close to the predicted value $\theta=0.667$. For $\mu=0.8$, we fit $P(x)$ by the functional form $P_0+C_0 x^{1-\mu}$ predicted in Eq.(\ref{case1}). The fit is very good as shown in  Fig.\ref{figure2}(b).

\begin{figure}[htb!]
\centering \includegraphics[width=.48\textwidth]{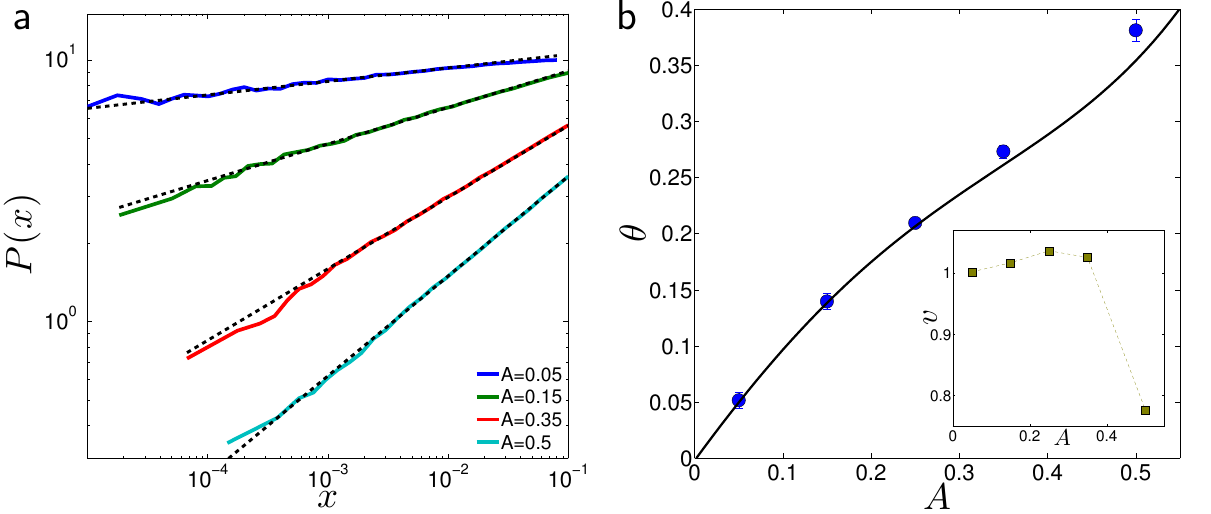} 
  \caption{(a) $P(x)$ for $\mu=1$, $N=512^2$ with different $A$. (b) Dependence of $\theta$  on $A$ for $\mu=1$. The blue dots are the numerical values extracted from (a), and the black line is the theoretical prediction Eq.(\ref{thetaonA}) using the 5 measured values of $v$ (see inset) and a $5th$ order polynominal interpolation to guide the eye.}\label{figure3}
\end{figure}

 For $\mu=1$, $\theta$ continuously depends on the kick amplitude, $A$ and the bias $v$. We plot the measured value of $\theta$ and the theoretical prediction in Fig.\ref{figure3}, and find once again a very good agreement.

 \begin{figure}[htb!]
\centering \includegraphics[width=.38\textwidth]{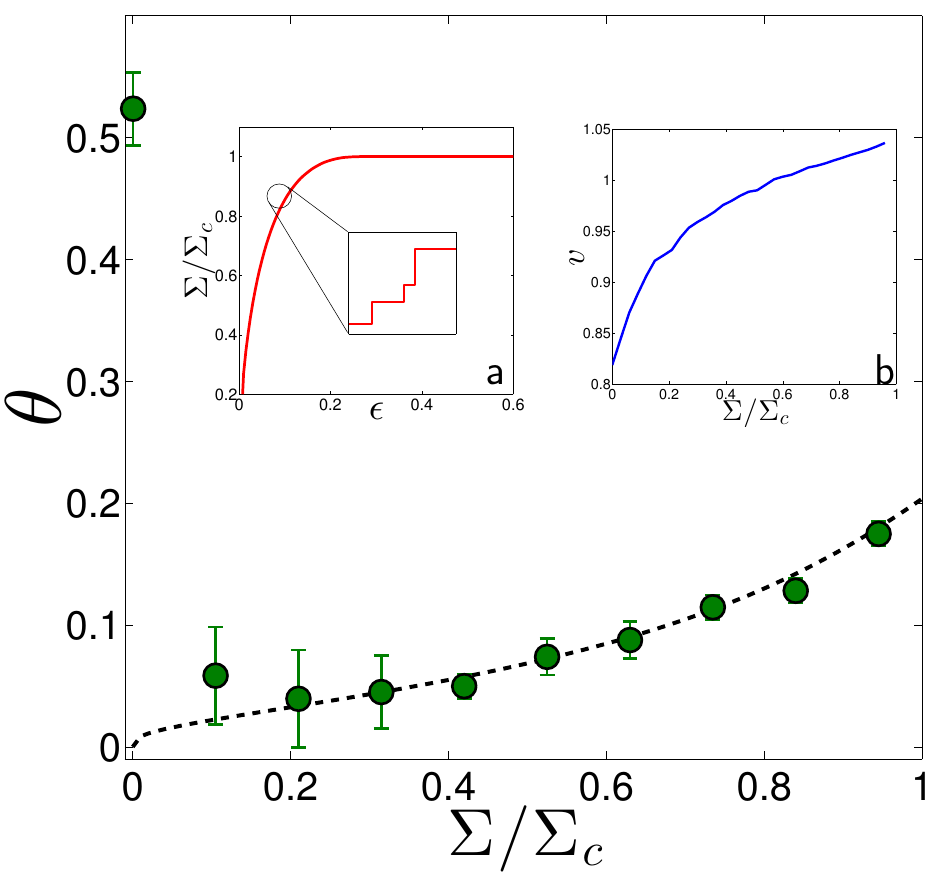} 
    \caption{$\theta$ as function of  the relative stress $\Sigma/\Sigma_{c}$, where $\Sigma_{c}$ is the yield stress for  $\mu=1$, $A=0.3$ and $N=1024^2$. Right after the quench,  $\theta=0.53\pm0.03$. As $\Sigma$ is increased, a sharp drop to a lower value occurs.  Inset (a): the stress {\it v.s.}  plastic strain curves, zooming in a finite size system reveals a staircase.  Inset (b): the bias $v$'s dependence on $\Sigma/\Sigma_{c}$, from which we get the black dashed line in the main panel as the theoretical prediction using Eq.\ref{thetatransient}.}\label{figure5}
\end{figure}

\section{\label{TB}Transient behavior}
Consider a liquid state with $\Sigma=0$. We model it as a configuration where many blocks are unstable, due to thermal fluctuations. The total initial distribution (including stable and unstable sites) $P_t(x)$ must be symmetric around $x=1$, and display tails in the unstable regions $x<0$ and $x>2$. Next, we suddenly quenched the system by setting the temperature $T$ to zero.  Importantly, the symmetry $\Sigma=0$ imposes that in the dynamics that follows, the same number of sites become unstable at $\sigma>1$ and $\sigma<-1$, implying that the drift $v=\langle \sigma_u\rangle = 1-\langle x_u\rangle =0$. According to Eq.(\ref{thetaonA}) we thus expect $\theta=1/2$ in our mean-field approximation. This prediction is consistent with the molecular dynamics simulations of \cite{Karmakar10a} which find  $\theta\approx 0.6$ after a quench both for $d=2$ and $d=3$. It  is tested numerically in our  model in Fig.\ref{figure5} where we find $\theta=0.53\pm0.03$ for an initial condition where $P_t(x)$ is uniform in $[-1,3]$. Numerically we found consistent results as long as enough unstable sites are initially present.
 
This situation dramatically changes however as soon as $\Sigma$ increases  from 0. Avalanches are then triggered, and the stress {\it v.s.} plastic strain curves (experiments generally report the stress vs the total strain, which is a plastic strain $\epsilon$ plus an elastic contribution $\Sigma/E$), although smooth in the thermodynamic limit, consists of steps as shown in inset (a) of Fig.\ref{figure5}\cite{Lin15}. Inside these avalanches (horizontal segment in inset), the stress is fixed and one can measure a drift $v$, as shown in inset (b) of Fig.\ref{figure5}. However, when the stress goes up in between avalanches (vertical segment in inset (a)),  all sites are shifted toward negative $x$, leading to an additional contribution to the drift. Its magnitude in the thermodynamic limit follows $d\Sigma/d\gamma=v d\Sigma/d\epsilon$. This contribution is large (and dominant) initially and vanishes at $\Sigma_c$ due to the shape of the stress-strain curves displayed in inset (a) of Fig.\ref{figure5}. 
Using Eq.(\ref{thetaonA}), we obtain the mean-field prediction for $\theta$ in a transient:
\begin{equation}
\theta=\frac{1}{\pi}\arctan(\frac{\pi A}{v(1+\frac{d\Sigma}{d\epsilon})})\label{thetatransient}
\end{equation}
This prediction is tested in Fig.\ref{figure5} and works remarkably well. Most importantly, Fig.\ref{figure5} is very similar to what is found in finite-dimensional elasto-plastic model \cite{Lin15}. This correspondence  indicates that our mean-field model correctly captures non-trivial effects present in finite dimensions, that were unexplained in the past.% {\red In a very recent molecular dynamics simulation work, by Hentschel, {\it et al}\cite{Hentschel15}, the authors investigate an athermal strained amorphous solids, and compute the pseudo-gap exponent via finite size effects\cite{Karmakar10a}, as a function of the imposed strain. They found that the exponent exhibits a sudden drop initially, followed by a slow increase as the strain increase, which is exactly what the mean field model predicts. We emphasize that this striking non-monotonic behavior observed in real systems, is nothing else but due to the fact that the interaction kernel decays as $1/r^{d}$ in real space, and a  L\'evy index $\mu=1$ for the corresponding random noise distribution. }

\section{\label{RSD}Role of spatial dimensions}

  \begin{figure}[htb!]
\centering \includegraphics[width=.49\textwidth]{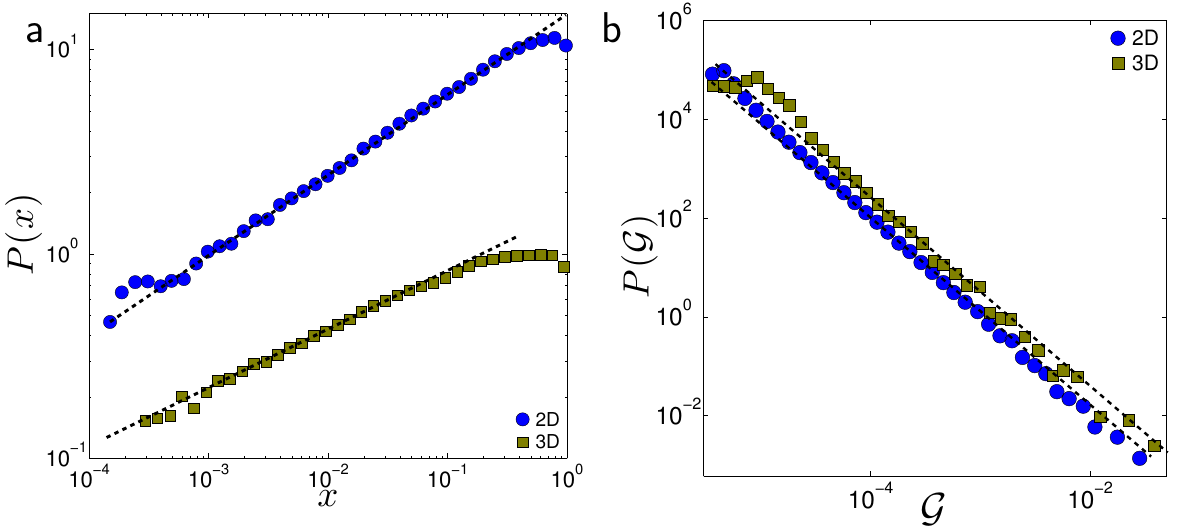} 
    \caption{(a) $P(x)$ at $\Sigma_c$ for the shuffled Eshelby kernel for $d=2$ and $d=3$. We find $\theta_{2D}^{SF}=0.39\pm0.02$ and $\theta_{3D}^{SF}=0.29\pm0.02$ (b) Direct measurements of the distribution $P({\cal G})$ of the Eshelby kernel corresponding to a stress drop unity at the origin, from which we extract $A_0=0.7\pm0.2$ (d=2), and $A_0=0.4\pm0.1$ (d=3).}\label{figure4}
\end{figure}
  \begin{figure}[htb!]
\centering \includegraphics[width=.4\textwidth]{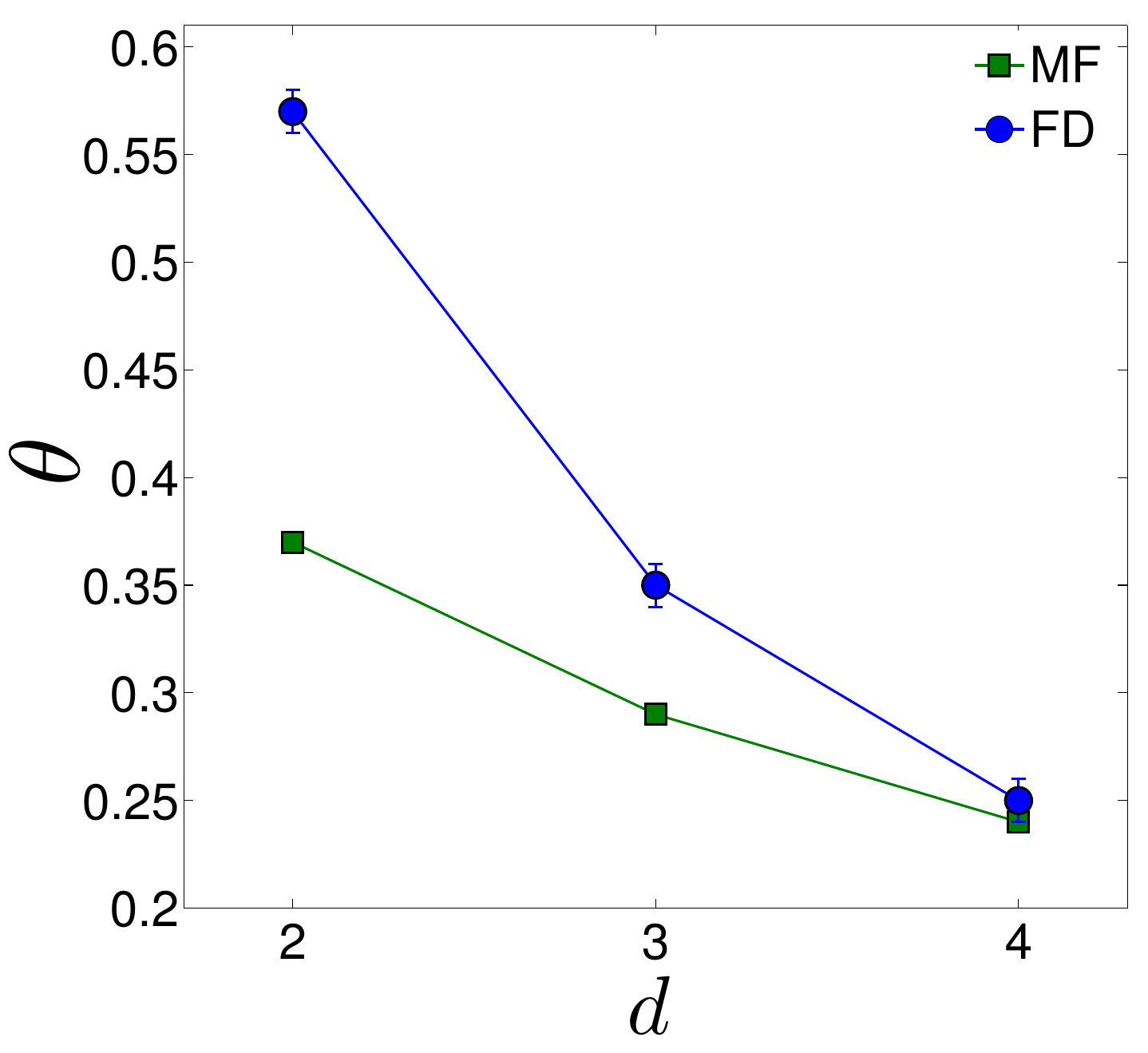} 
    \caption{Comparison between the measurements of the pseudo-gap exponent $\theta$  in finite dimensional elasto-plastic models (FD, circle) and its mean field prediction (MF, square) as a function of the spatial dimension $d$. As $d$ increases, the difference become smaller and undetectable for $d=4$.}\label{thetaond}
\end{figure}
In finite dimension the interaction kernel ${\cal G}$ is well-described by the Eshelby kernel \cite{Picard04,Kabla03, Desmond13}. For $d=2$ for example it follows $\mathcal{G}_{ji}(\vec{r}_i-\vec{r}_j)\sim\cos(4\phi)/|\vec{r}_i-\vec{r}_j|^2$ where $\phi$ is the angle between the shear direction and $\vec{r}_i-\vec{r}_j$. To quantify finite-dimensional effects, we consider the mean-field model obtained by shuffling $\mathcal{G}_{ji}$ randomly at each event:  
 \begin{equation}
\label{05}
x_j(l+1)=x_j(l)+\mathcal{G}_{k_l(j)i}(\vec{r}_i-\vec{r}_j) \sigma_i (l)
\end{equation}
where $k_l(j)$ is a random permutation of all indices $j\neq i$. We then measure $\theta$ at $\Sigma_c$, considering 
both the two-dimensional and three-dimensional Eshelby kernel, and show our results in Fig.\ref{figure4}.a.

We find that $\theta_{2D}^{SF}=0.39\pm0.02$ and $\theta_{3D}^{SF}=0.29\pm0.02$. These values are in very good agreement with the prediction Eq.(\ref{thetaonA}), $\theta_{2D}^{MF}=0.37\pm0.05$, $\theta_{3D}^{MF}=0.29\pm0.04$. To compute these predictions we first extracted the pre-factor  $A_0$ characterizing the amplitude of the power law distribution of the spatial kernel, as done in Fig.\ref{figure4}.b.  Following Eq.(\ref{05}), the noise amplitude $A$ is related to  $A_0$ as $A=A_0\langle |\sigma_u|\rangle$, where $\langle |\sigma_u|\rangle$ is the mean absolute stress of unstable sites when they relax. The ratio $A/v$ which determines $\theta$ follows  $A/v=A_0\langle |\sigma_u|\rangle/\langle \sigma_u\rangle $. For the mean field model with shuffled kernel, we numerically find that $\langle |\sigma_u|\rangle/\langle \sigma_u\rangle \approx 1$, ($\langle |\sigma_u|\rangle/\langle \sigma_u\rangle=1.01\pm 5\times 10^{-4}$ in $d=2$, $1.0\pm 2\times 10^{-5}$ in $d=3$), a result that must hold when most sites become unstable in the direction of the shear, {\it i.e.}, at the boundary $x<0$. So the result further simplifies to $A/v\approx A_0$.

These mean-field values for $\theta$ are systematically smaller than observations in finite dimensions where $\theta_{2D}\approx 0.57$ and $\theta_{3D}\approx 0.35$ \cite{Lin14,Salerno13}. To complete this comparison, we computed the mean field predictions and the finite dimensional values of $\theta$ in $d=4$ in Appendix.C. Those results are summarized for different spatial dimensions  in Fig.\ref{thetaond}.  We observe that the difference between mean-field prediction and finite-dimensional observations become smaller as the spatial dimension increases, and becomes indistinguishable  numerically for $d=4$. Our work is thus consistent with the critical dimension being $d_c=4$. There is currently no Ginzburg-type argument to justify why this would be the case.

%
% We shuffle $\mathcal{G}_{ji}$, by rearranging the index $i$,$j$ randomly at each event, which is called the shuffled Eshelby kernel, and compare $\theta$ at $\Sigma_c$ with the raw one. Direct measurement of $w(\xi)$ for the Eshelby kernel yields $\theta_{2D}^{MF}=0.37$, and $\theta_{3D}^{MF}=0.29$, if $v=1$(which we verify in the numerics), consistent with the direct measurement of $\theta$ for the shuffled case, shown in Fig.\ref{figure4}. We find the spatial correlation effects between successive events at 2D and 3D still affects the universality, and we obtain different $\theta$ exponents from the mean field results, $\theta_{2D}=0.61\pm 0.04$, $\theta_{3D}=0.37\pm 0.02$. Nevertheless, the difference between the mean field $\theta$ and spatial $\theta$ decreases as the dimension increases, which suggests that they may conincide each other above some critical dimensions $d_c$. Moreover, the mean field result capture the fact that $\theta$ decreases as the dimension increases, as observed in the spatial model, from which, we may expect in both cases, $\theta\rightarrow 0$ in the $d\rightarrow \infty$ limit.
%

\section{\label{CO}Conclusion}
We have analytically solved a mean-field model of the plasticity in amorphous solids, focusing on the exponent $\theta$ characterizing the density of shear transformations and the stability toward avalanches. The most surprising result is that $\theta$ is found to be universal after an isotropic quench, but to be otherwise stress-dependent and non-universal. Two pieces of evidence supports that our model is the correct mean-field description of elasto-plasticity, for which plasticity is governed by local rearrangements interacting elastically.  First, we make the surprising prediction that $\theta$ varies non-monotonically with the stress level, a fact previously observed  \cite{Lin15} but unexplained even at a qualitative level.  Second, our prediction for $\theta$  becomes accurate as the spatial dimension increases. Most importantly, our predictions are close to observations in three dimensions. In four dimensions predictions and observations cannot be distinguished, suggesting that the critical dimension is $d_c=4$. 

Overall, our work is consistent with the notion that the yielding transition at $\Sigma_c$ is a dynamical phase transition, but supports that it is a transition of a curious kind, where exponents can depend continuously on parameters. It is still unclear if exponents in finite dimensions can be computed via a perturbation around some critical dimension, where the mean-field solution becomes exact  but non-universal. A first step in that direction would be to build a Ginzburg-type criterion to predict the critical dimension. 

It is interesting to reflect on  how predictive the value of $\theta$ measured in elasto-plastic models should be to describe real materials at the yield stress.  From the present analysis this is a priori not obvious at all, since exponents are not expected to be universal (except at zero shear) and to potentially depend on the details of the model.   However, measurements in molecular dynamics simulations and elasto-plastic models appear to yield very similar values for $\theta$ \cite{Lin14,Salerno13}. A possible explanation is that in elasto-plastic models, as long as most sites become unstable  along the direction of the imposed shear (and not opposite to it), we predict  $\theta$ to be only a function of the coefficient $A_0$ characterizing the Eshelby kernel ${\cal G}$, as discussed in Section VII. The similarity between elasto-plastic models and molecular dynamics simulations may thus reflect the accuracy of the Eshelby kernel in capturing the interaction between shear transformation in real amorphous solids. 

 Although we have focused on amorphous solids, it is very plausible that this model applies to disordered crystals as well, where plasticity is mediated by dislocations whose motions interact with the same Eshelby kernel studied here. It will thus be very interesting to test our predictions in both classes of materials. 
 
Finally, the concept of marginality has been very influential in electron glasses \cite{Efros75} but its validity is still debated in that context \cite{Muller14,Palassini12}. Introducing dynamical mean-field models of the type discussed here may resolve this question.

\section*{ACKNOWLEDGMENTS}
It is a pleasure to thank J.P. Bouchaud, E. DeGiuli, M. Fall, T. Gueudre, H. Hajaiej, E. Lerner, M. Muller, A. Rosso, A. Vasseur, L. Yan for discussions related to this work, and reviewers for constructive comments. MW acknowledges support from NSF CBET Grant 1236378 and MRSEC Program of the NSF DMR-0820341 for partial funding. 

\renewcommand{\theequation}{A.\arabic{equation}}
\setcounter{equation}{0}

\renewcommand{\thefigure}{A.\arabic{figure}}
\setcounter{figure}{0}
\section*{APPENDIX A: Probabilistic interpretation }
We now present a probabilistic interpretation of these results. The dynamics of a single block is equivalent to a random walker with  L\'evy index $\mu$ in $x$ space with a bias $-v$ towards the absorbing boundary at $x=0$.   We introduce the Hurst exponent for the mean absolute displacements without  bias
\begin{equation}
\langle |x-x_0|\rangle \sim \gamma^{H}
\end{equation}
 for L\'evy flights $H=1/\mu$ if $\mu<2$, and $H=1/2$ if $\mu\geq 2$. We define $P_+(x,x_0,-v,\gamma)$ as the probability to find the walker starting from $x_0$ with a bias $-v$ at position $x$ after a time $\gamma$, without having crossed the $x<0$ half-line (in our argument the presence of another wall at $x>2$ is irrelevant for the small $x$ behavior of interest here).
A block that has just relaxed starts from $x_0=1$, and reaches $x$ after some time $\gamma$. In the stationary state, we obtain
\begin{equation}
P(x)=\int_{0}^{\infty} P_+(x,1,-v,\gamma) d\gamma
\end{equation}
For small $x$, this integral is dominated by $\gamma \simeq 1$. Indeed because $v\simeq 1$, thus for $\gamma \ll 1$ the walker is still far from the origin and has negligible chance to hit the origin due to the random noise. Instead, for $\gamma \gg 1$ the walker is very likely to have entered the unstable region due to the bias. Thus we have for small $x$:
\begin{equation}
P(x)\sim  P_+(x,1,-v,1)\label{Px}
\end{equation}

For L\'evy flight with no bias, it is known that the absorbing boundary leads to $P_+(x,x_0,0,\gamma)\sim x^{\theta}$ at small $x$ \cite{Zumofen95,Zoia07}. We define the persistence probability $S(x_0,\gamma)$ as the probability for the walker starting at $x_0>0$ to have remained in the positive axis at time $\gamma$. In the limit $x_0\ll \gamma^{H}$, $S(x_0,\gamma)\sim \gamma^{-\kappa}$, and $\kappa$ is called the persistence exponent. There is a general scaling relation in the absence of bias, $\theta=\kappa/H$\cite{Zoia09}. In addition, for L\'evy-flights according to the Sparre Andersen theorem\cite{Andersen1954} $\kappa=\frac{1}{2}$, so for a random walker without bias ($v=0$),  $\theta=\mu/2$ for $\mu<2$ and $\theta=1$ for $\mu\geq 2$. 

We now extend these results to the case where there is a bias. Time-reversal symmetry implies:
\begin{equation}
\label{009}
P_+(x,x_0,-v,\gamma)=P_+(x_0,x,v,\gamma)
\end{equation}
Thus we transfer the problem to the backward process, where $x$ is now the starting position, with a drift away from the absorbing boundary. We expect $P_+(x_0,x,v,\gamma)$ to have the scaling form:
 \begin{equation}
 \label{007}
 P_+(x_0,x,v,\gamma)=\frac{1}{\gamma^{H^{\prime}}} F(\frac{x_0}{\gamma^H}, \frac{x}{ \gamma^H}, \frac{v} {\gamma^{H-1}} )
 \end{equation}
Integrating over $x_0$, we obtain the persistence probability $S(x,v,\gamma)=\gamma^{H-H^{\prime}} G(\frac{x}{\gamma^H}, \frac{v}{\gamma^{H-1}})$ where $G$ is some function. Because $S(x, v, 0)=1$, we must have $H^{\prime}=H$, therefore 
\begin{equation}
\label{10}
 S(x,v,\gamma)= G(\frac{x}{ \gamma^H}, \frac{v }{ \gamma^{H-1}})
\end{equation}
We define the normalized surviving probability density as 
\begin{equation}
p_{x,v}(x_0,\gamma)=\frac{P_+(x_0,x,v,\gamma)}{\int_{0}^{\infty} P_+(x_0,x,v,\gamma)dx_0}=\frac{P_+(x_0,x,v,\gamma)}{S(x,v,\gamma)}
\end{equation}
from which we get, together with Eq.(\ref{007}):
\begin{equation}
F(y_0,y,\omega)= G(y,\omega) p_{y,\omega}(y_0)
\end{equation}
where we use the scaled variable $y=x/\gamma^{H}$, $\omega= v/\gamma^{H-1}$, and $p_{y,\omega}(y_0)=p_{x,v}(x_0,\gamma)\gamma^H$.  It is clear from its definition that $p_{y,\omega}(y_0)$ must converge to a constant  $p_{0,\omega}(y_0)$ as $y\rightarrow 0$. Thus the asymptotic behavior of $F$ in the small $y$ limit is that of the surviving probability $G(y,\omega)$.  From this and Eqs.(\ref{Px},\ref{009},\ref{007}) we get the central result that $P(x)\sim S(x,v,\gamma=1)$. In other words, $P(x)$ is related to the survival of a walker starting at $x$ with a positive bias $v$ after a time of order unity. According to Eq.(\ref{10}), we have then: $P(x)\sim G(x, v)$, where $v$ is a constant of order one.
%We now make use of:
% \begin{equation}
%S(x,\gamma)=G(y,\omega)=G(\frac{x}{\gamma^{H}}, \gamma^{1-H})
%\end{equation} 

 \textbf{ Case I: $1<\mu<2$}. In this case, $ 1/2<H<1$. From Eq.(\ref{10}) we get that for $y=x/\gamma^H$ constant and $\gamma\ll 1$, $S(x,v,\gamma)=G(y,0)$, i.e. the drift term $v$ is irrelevant at small times. In the absence of bias it is known that $G(y,0)\sim y^{\mu/2}$ \cite{Zumofen95}. As $\gamma$ increases toward one, the effect of the bias becomes of the order of the noise, we thus expect it to only affect the survival probability by a numerical prefactor, implying that the result:
 \begin{equation}
\theta=\frac{\mu}{2}
\end{equation}
holds even with a finite bias, as proven above. 

 \textbf{ Case II: $\mu<1$}. In this case, $H>1$, and the bias is relevant at small times according to Eq.(\ref{10}). In that limit, $\gamma\ll 1$, we thus get $S(x,v,\gamma)=G(y,\infty)=1$. Once again when $\gamma$ increases and becomes of order one, the effect of the noise becomes of order of the bias, and will affect the survival probability by some numerical prefactor.  We thus get $G(y,1)\sim 1$ for small $y$, leading to $\theta=0$ as derived above.

 \textbf{ Case III: $\mu=1$}. In this  case, $H=1$, and the velocity is marginal in Eq.(\ref{10}), implying that $S(x,v,\gamma)= G(x/\gamma^H, v )$. As first derived in\cite{Doussal09}, and reviewed in \cite{Bray13}, it is known that at long times, for $\mu=1$  the persistence exponent follows:
\begin{equation}
\kappa=\frac{1}{2}-\frac{1}{\pi}\arctan(C)
\end{equation} 
 where $C=v/l_{\mu}$, where in our notation $l_1=\pi A$, leading to $C=v/\pi A$. Thus $G(x/\gamma^H, v )\sim \gamma^{-\kappa}$ in that limit, which is only possible if $G(x/\gamma^H, v )\sim x^{\kappa/H}\sim x^\kappa$. 
 This corresponds to $\theta=\kappa$. After some manipulations it leads to:
 \begin{equation}
 \theta=\frac{1}{\pi}\arctan(\pi A/v)
 \end{equation}
as derived above.

\begin{figure}[thb!]
   \includegraphics[width=.4\textwidth]{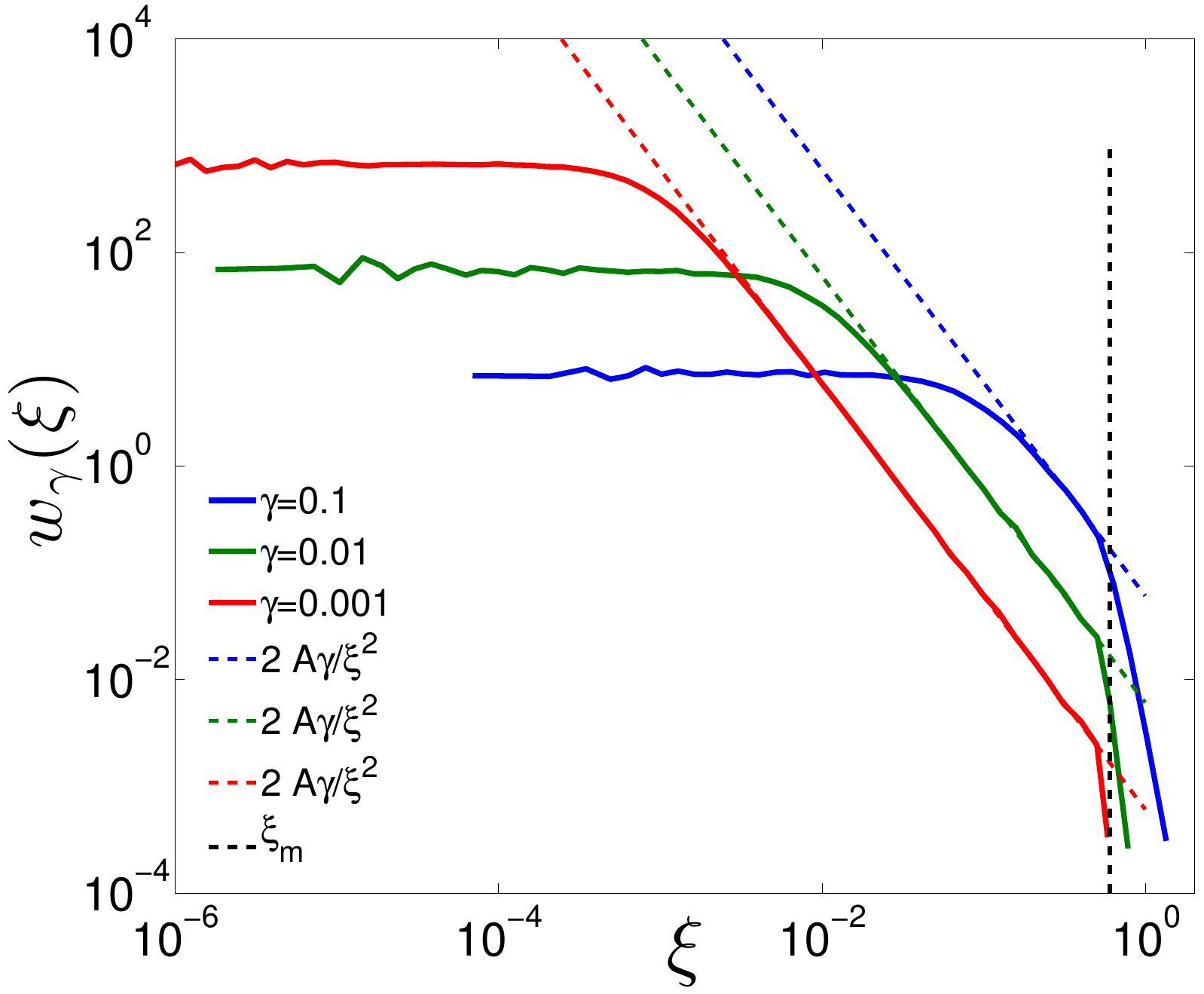} 
    \caption{Numerical calculation of $w_{\gamma}(\xi)$ at $\gamma=0.1, 0.01, 0.001$, $\mu=1$. The colorful dashed lines are the theoretical prediction, for $\gamma^{1/\mu}\ll \xi \ll \xi_m$. The black dashed line is at $\xi_m$, where the upper cut-off begins to play role.}\label{figureap1}
\end{figure}

\begin{figure*}[htb!]
   \includegraphics[width=.9\textwidth]{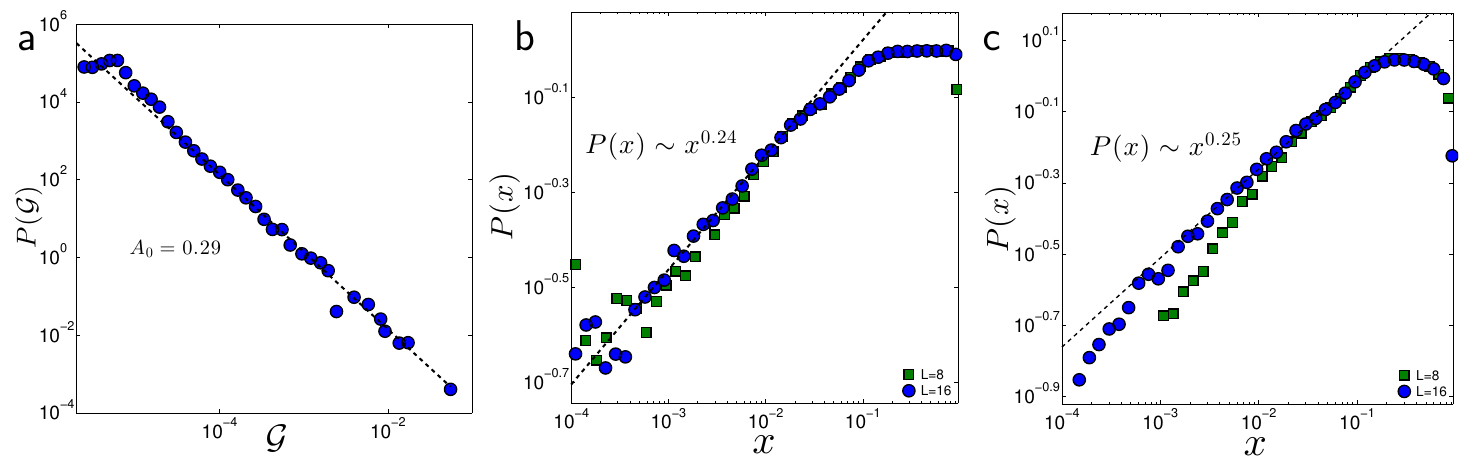} 
    \caption{(a) Direct measurements of $A_0$  as extracted from the distribution $P{\cal G})$ of the Eshelby kernel in $d=4$. We extract $A_0=0.29\pm0.4$  and predict $\theta_{4D}^{MF}=0.24\pm0.04$. (b) $P(x)$ at $\Sigma_c$ using extremal dynamics for the shuffled Eshelby kernel for $d=4$, where $\theta_{4D}^{SF}=0.24\pm0.01$. (c) Finite dimensions measurement of $P(x)$  in $d=4$ at fixed stress $\Sigma_c=0.5025$, where $\theta_{4D}=0.25\pm0.01$.}\label{figureap2}
\end{figure*}

\section*{APPENDIX B: Asymptotic behavior of $w_{\gamma}(\xi)$}
The probability distribution of $\xi$ is 
\begin{equation}
w(\xi)=\frac{A}{N}  |\xi|^{-\mu-1}
\end{equation}
with the lower cut-off at $|\xi|_{c}=(\frac{2 A}{\mu})^{1/\mu} N^{-\frac{1}{\mu}}$, and the upper cut-off $|\xi|_{m}=(\frac{2A}{\mu})^{\frac{1}{\mu}}$. The corresponding Fourier transformation is 
\begin{align}
1-w(k)=\frac{2A}{N}\int _{\xi_{c}}^{\xi_{m}} \frac{1}{\xi^{\mu+1}} (1-\cos( k \xi)) d\xi \\
=\frac{2A}{N} |k|^{\mu}  \int_{k \xi_{c}}^{k \xi_{m}}  \frac{1}{y^{\mu+1}} (1-\cos(y)) dy \nonumber
\end{align}
in the limit $N\rightarrow \infty$, the above integral becomes
\begin{align}
1-w(k) &= \frac{2A}{N} |k|^{\mu}  \int_{0}^{k\xi_m}   \frac{1}{y^{\mu+1}} (1-\cos(y)) dy\nonumber \\
&=\frac{2AI_{\mu}}{N} |k|^{\mu}H(k\xi_m)
\end{align}
where $I_{\mu}=-\cos(\frac{\mu\pi}{2})\Gamma(-\mu)$, and
\begin{equation}
 H(y)=\frac{1}{I_{\mu}}\int_0^{y} \frac{1-\cos(t)}{t^{\mu+1}} dt 
 \end{equation}
which behave as
\begin{align}
H(y) &= 1  &      y \gg 1   \nonumber \\
H(y) &= B_0 y^{2-\mu}   & y\ll 1 \label{Hy} 
\end{align}
where $B_0= \frac{1}{2(2-\mu)I_{\mu}} $.
 In the large $N$ limit, the coarse-grained distribution $w_{\gamma}(k)$ becomes
\begin{align}
w_{\gamma}(k)=(1-\frac{2AI_{\mu}}{N}|k|^{\mu}H(k\xi_m))^{\gamma N} \nonumber\\
\rightarrow \exp\{-\gamma(l_{\mu} |k|)^{\mu}H(k\xi_m)\}
\end{align}
where $l_{\mu}=(2AI_{\mu})^{1/\mu}$. We are interested in $w_{\gamma}(\xi)$ in the limit $\gamma\rightarrow 0$, so it is convenient to define 
\begin{align}
\tilde{\xi}     &=\frac{\xi}{l_{\mu}\gamma^{1/\mu}}   \nonumber \\
\tilde{\xi}&_m=\frac{\xi_m}{l_{\mu}\gamma^{1/\mu}}   \label{lmu}
\end{align}
 Making use of Eq.(\ref{Hy}), we can decompose $w_{\gamma}(\xi)$ as
\begin{align}
&w_{\gamma}(\xi)\approx \frac{1}{\pi l_{\mu}\gamma^{1/\mu}} \int_{0}^{\infty} \exp(-y^{\mu})\cos(y\tilde{\xi}) dy  \nonumber\\
&+ \frac{1}{\pi l_{\mu}\gamma^{1/\mu}} \int_{0}^{\tilde{\xi}_m^{-1}} \{e^{-y^{\mu} H(y\tilde{\xi}_m)}-e^{-y^{\mu}} \}\cos(y\tilde{\xi}) dy  \nonumber \\
&= w_1+w_2
\end{align}

Here $w_1=\frac{1}{l_{\mu} \gamma^{1/\mu}} \mathcal{L}_{\mu} (\tilde{\xi})$, and in the limit we are interested in $\gamma\rightarrow0$, it reduces to
\begin{equation}
w_1=\frac{\gamma A }{\xi^{\mu+1}}\label{J1}
\end{equation}
We discuss the $w_2$ contribution in the two regimes:\\

\textbf{Case I: $1\ll \tilde{\xi} \ll \tilde{\xi}_m$}.
In this regime, we can expand the exponential terms to the first order, and obtain
\begin{align}
w_2&\approx \frac{1}{\pi l_{\mu} \gamma^{1/\mu}} \int_{0}^{\tilde{\xi}_m^{-1}} (y^{\mu}-B_0\tilde{\xi}_{m}^{2-\mu} y^2) dy \nonumber\\
&=\frac{1}{\pi l_{\mu }\gamma^{1/\mu}} \frac{2-\mu}{3(\mu+1)} \tilde{\xi}_m^{-\mu-1} \nonumber\\
&\sim \frac{\gamma }{\xi_{m} ^{\mu+1}}
\end{align}
which turns out to be much smaller than Eq.(\ref{J1}) and therefore negligible in this regime!

\textbf{Case II: $1\ll \tilde{\xi}_m \ll \tilde{\xi}$}.
 In this case, we can extends the integral to infinity, because of the oscillating factor,
\begin{align}
w_2&\approx\frac{1}{\pi l_{\mu }\gamma^{1/\mu}} \int_{-\infty}^{\infty} \{\exp(-B_0 \tilde{\xi}_m^{2-\mu} y^2+iy\tilde{\xi})\nonumber \\
&-\exp(-|y|^{\mu} + iy\tilde{\xi})\} dy
\end{align}
where the second term cancels $w_1$, and the resulting distribution $w_{\gamma}(\xi)$ in this limit is
\begin{align}
w_{\gamma}(\xi)&\approx \frac{1}{l_{\mu} \gamma^{1/\mu}}  \frac{1}{\sqrt{\pi B_0 \tilde{\xi}_m^{2-\mu} }} \exp(-\frac{\tilde{\xi}^2 } {4B_0 \tilde{\xi}_m^{2-\mu}})\nonumber \\
&\sim \gamma^{-1/2}  \exp(-\frac{\xi^2} {4\gamma B_0 l_{\mu}^{\mu} \xi_m^{2-\mu}})
\end{align}
which is much smaller than Eq.(\ref{J1}) at $\xi\approx \xi_{m}$, and behaves as an upper cut-off in $w_{\gamma}(\xi)$. Our numerical calculation of $w_{\gamma}(\xi)$ is consistent with the above theoretical analysis, shown in Fig.\ref{figureap1}.

\section*{APPENDIX C: $P(x)$ in $d=4$}
To confirm that the validity of mean field as the dimension increases, we simulate the elasto-plastic model in $d=4$. Direct measurement of the coefficient $A_0$ characterizing the distribution $P({\cal G})$ yields $A_0=0.29\pm0.05$, and numerically we find $\langle |\sigma_u|\rangle/\langle \sigma_u\rangle=1$ without any unstable sites hitting the other absorbing boundary, from which one predict $\theta_{4D}^{MF}=0.24\pm0.02$, shown Fig.\ref{figureap2}(a). For the shuffled kernel, we measure $\theta_{4D}^{SF}=0.24\pm0.01$, Fig.\ref{figureap2}(b), consistent with the theoretical prediction. The finite dimensional measurement yields $\theta_{4D}=0.25\pm0.01$, as shown in Fig.\ref{figureap2}(c).

\bibliography{Wyartbibnew}

\end{document}